\documentclass[11pt,osid,twoside,openright]{article}
\usepackage{graphics}
\usepackage{epsfig}
\usepackage{mathcomp}
\usepackage{amsmath}
\usepackage{amsfonts}
\usepackage{amssymb}
\usepackage{graphics}
\usepackage{epsfig}
\usepackage{amsmath}
\usepackage{osid}                         %

\title{Two-way Protocol for Quantum Cryptography with Imperfect Devices}
\author{Marco Lucamarini\thanks{marco.lucamarini@unicam.it},
        Alessandro Cer\'e,
        Giovanni Di Giuseppe,
        Stefano Mancini,
        David Vitali, and
        Paolo Tombesi
        \\{\footnotesize\it
        Dipartimento di Fisica, Universit\'a di Camerino, I-62032 Camerino,
        Italy}\\[2ex]
        }

\begin{document}

\maketitle

\begin{abstract}
The security of a deterministic quantum scheme for communication,
namely the LM05~\cite{Luc05}, is studied in presence of a lossy
channel under the assumption of imperfect generation and detection
of single photons. It is shown that the scheme allows for a rate
of distillable secure bits higher than that pertaining to
BB84~\cite{BB84}. We report on a first implementation of LM05 with
weak pulses.
\end{abstract}

%----------------------------------------------
%----------------------------------------------
\section{Introduction}

Deterministic quantum schemes (DQS) for secure communication have
recently gained interest and diffusion in the field of quantum
cryptography \cite{DQS,Luc05}, and their first experimental
`proofs of principle' have been already
completed~\cite{ExpDQS,Cere06}. Despite the main achievement of a
\textit{secure direct communication} (A. Beige \textit{et al.} in
\cite{DQS}~) is still quite far, DQS can provide better security
and higher transmission rates in the quantum key distribution
(QKD) process than traditional schemes like the BB84~\cite{BB84}.

One of these DQS, namely the LM05~\cite{Luc05}, saturates the
Holevo bound for QKD~\cite{Cab00}, and is quite practical to
implement as it does not require entanglement to
work~\cite{Cere06}. The security of LM05 against eavesdropping in
case the users (Alice and Bob) are endowed with a perfect
equipment was discussed in~\cite{Luc05}. Specifically LM05 results
robust against a general eavesdropping on a noisy but lossless
channel, and explicit thresholds were given in case of individual
attacks by the eavesdropper (Eve); furthermore a particular
strategy by Eve on a noisy and lossy channel as described
in~\cite{Woj03} was also deemed as detectable by legitimate users.

In this work we relax the hypothesis of perfect equipment for
Alice and Bob. We take as photon source an attenuated laser that
produces weak pulses; these pulses can accidentally (and
uncontrollably) contain more than one photon. Furthermore Bob's
detectors are avalanche photodiodes (APD) that either `click' or
`not click', without counting the exact number of photons in the
pulse, and have nonunitary quantum efficiency and nonzero dark
counts probability. It has been shown that the conjunction of
imperfect devices with a lossy channel jeopardizes the security of
QKD~\cite{Cur04}. The main threat is represented by a
photon-number splitting attack (PNS) in which an almighty Eve
exploits the multiphoton pulses to acquire information, whilst
concealing her presence behind the expected
losses-rate~\cite{Ben92b,Hut95,Yue96}. PNS attacks are currently
the main limitation to a long-distance BB84 realized with weak
pulses~\cite{Lut00,Bra00}.

The paper is organized as follows. In section I we review the LM05
protocol and describe the PNS attacks against it. In section II we
theoretically study the security against these attacks in terms of
the rate of distillable secure bits. In section III we describe
the first experimental test of LM05 with weak pulses.

\section{Theory}

The LM05 protocol works as follows~\cite{Luc05}. Bob prepares a
photon in one of the four polarization states $|0\rangle$,
$|1\rangle$, $|\pm\rangle=1/\sqrt{2}\left(
|0\rangle\pm|1\rangle\right)  $, with $|0\rangle$, $|1\rangle$
eigenstates of the Pauli operator $\widehat{\sigma}_{z}$, and
sends it to Alice. With probability $c$ Alice measures the photon
(\textit{control mode, }CM) as she would do in the BB84 protocol.
This guarantees that the scheme is at least as secure as the BB84.
Otherwise, with probability $1-c$, she uses the photon to encode a
bit (\textit{message mode, }MM) by flipping (logical value `$1$')
or not flipping (logical value `$0$') it. After that she sends the
photon back to Bob. To flip the photon without knowing its state
Alice uses the operation $i\widehat{\sigma}_{y}$, that acts as a
universal `equatorial NOT' gate~\cite{Gis02}. Bob can
deterministically decode Alice's message by measuring the qubit in
the same basis he prepared it, without demand for a classical
channel. We point out that LM05 does not allow for a direct
communication when the channel is noisy or lossy. As explained
in~\cite{Hof05} it is not possible so far to achieve both a
reliable and secure delivery of a message: if one uses the error
correction protocol~\cite{Ben92c,Bra94} to make the communication
reliable Eve can capture a non negligible amount of information,
while if one uses the privacy amplification protocol~\cite{Ben95}
to make the communication secure Bob has no means to reconstruct
Alice's original message. Whether a secure and reliable direct
communication in presence of noise or losses is really possible is
still an open question.

In the following we describe two PNS attacks, that provide full
information to Eve while remaining completely undetected.
Notwithstanding the analysis includes also strategies of the same
kind that are only partially informative to Eve: with privacy
amplification~\cite{Ben95} Alice and Bob can remove any remaining
information from Eve, according to what explained
in~\cite{Lut00}.\\
\indent When the photon source is a laser attenuated with an
average photon number per pulse $\mu$, the probability to have $n$
photons in a single pulse is given by~\cite{foot1}:
\begin{equation}
P_{n}\left(  \mu\right)  =\frac{\mu^{n}}{n!}e^{-\mu}.\label{Pn(miu)}%
\end{equation}
A typical value used for $\mu$\ in the experiments is $0.1$ that
gives $P_{0}\simeq\allowbreak9\cdot10^{-1}$,
$P_{1}\simeq\allowbreak9\cdot10^{-2}$,
$P_{2}\simeq\allowbreak4.5\cdot10^{-2}$, and so on. This means
that with a
probability $P_{n}\left(  \mu\right)  $\ Bob prepares the state%
\begin{equation}
    \left\vert \psi\right\rangle ^{\otimes n}=
        \begin{array}[c]{c}
            \underbrace{\left\vert \psi\right\rangle
            \otimes...\otimes\left\vert \psi\right\rangle }\\
            n\,\,\rm{ times}
        \end{array}\label{Psi_n}
\end{equation}
rather than the desired state $\vert \psi \rangle $ ($\psi$
indicates one of the four polarizations of the photon prepared by
Bob).

It is known (Lutkenhaus's suggestion in \cite{Sca04}) that when
$n=3$ it exists a measurement $\mathcal{M}$ that provides a
conclusive result about the absolute polarization $\psi$ with
(optimal) probability $1/2$. Eve can exploit this fact to
eavesdrop on LM05 protocol in the following way. She performs a
quantum nondemolition measurement (QND) on the pulses as soon as
they exit Bob's station; this can be done without perturbing the
polarization $\psi$. When she finds $n<3$ she blocks the pulses.
On the pulses with at least three photons she executes
$\mathcal{M}$ and if the outcome is not conclusive she blocks
these pulses as well. When $n\geq3$ \textit{and} the outcome\ of
$\mathcal{M}$ is conclusive she prepares a new photon in the right
state $\psi$ and forwards it to Alice. Until here this attack is
completely analogous to the `IRUD-attack' described in
\cite{Sca04}. The only variant is that Eve waits for Alice
encoding and measures again the photon on the backward trip, to
know whether it has been flipped (in this case she finds the
orthogonal state $\left\vert \psi^{\perp}\right\rangle $) or not
(she finds $\left\vert \psi\right\rangle $). Since Eve did know
$\psi$, she can extract Alice's information without perturbing the
state. After that she forwards the photon in the correct state to
Bob. We call this first attack PNS$_{\mathcal{M}}$.

A second attack is more peculiar to LM05. This time suppose that
$n=2$ and call the two photons in the pulse $p_{1}$ and $p_{2}$.
As before Eve can know the number of photons per pulse through a
QND measure. When $n<2$ Eve blocks the pulses. When $n=2$ she
stores $p_{1}$ and forwards $p_{2}$ to Alice; this let her remain
undetected during a possible CM on the forward path. On the way
back Eve captures again $p_{2}$. To gain Alice's information she
must decide whether the polarizations of $p_{1}$ and $p_{2}$ are
parallel or antiparallel: in the first case she would deduce the
logical value `$0$'; in the second case she would deduce `$1$'.
But the discrimination between parallel and antiparallel spins is
not as simple as it appears at a first glimpse: while the
parallel-spin-state $\vert P\rangle =\vert \psi\rangle
_{p_{1}}\vert \psi\rangle _{p_{2}}$ is symmetric, the
antiparallel-spin-state $\vert AP\rangle =\vert \psi \rangle
_{p_{1}}\vert \psi^{\perp}\rangle _{p_{2}}$ is neither symmetric
nor antisymmetric. Upon symmetrizing $\vert AP\rangle $ we can
realize that it is not orthogonal to $\vert P\rangle $, and by
consequence it is not perfectly distinguishable from it (we remand
to \cite{Bar04}, \cite{Pry05} for a complete treatment of this
problem). Actually an optimal measurement $\mathcal{M^\prime}$ is
a nonlocal one and gives Eve a conclusive result (between $\vert
P\rangle $ and $\vert AP\rangle $) with a probability $1/4$
\cite{Bar04}. Hence Eve can block all the `inconclusive' pulses to
gain full information and still remain undetected. However it
remains open the question of which photon must Eve forward to Bob.
The measurement $\mathcal{M^\prime}$ consists in a generalized
measurement with projectors in the four-dimensional Hilbert space
given by: $\Pi_{A}^{p_{1},p_{2}}=[\vert \psi\rangle\vert
\psi^{\perp}\rangle-\vert \psi^{\perp}\rangle\vert \psi\rangle]
    \otimes[ \langle \psi\vert \langle
\psi^{\perp}\vert -\langle \psi ^{\perp}\vert \langle
\psi\vert]/2$ and $\Pi_{S}^{p_{1},p_{2}}
=I_{4}-\Pi_{A}^{p_{1},p_{2}}$.
The conclusive answer is related to the antisymmetric state $\Pi_{A}%
^{p_{1},p_{2}}$. Upon obtaining this result Eve does not know
whether to give Bob the state $\vert \psi\rangle $ or the state
$\vert \psi^{\perp}\rangle $, because she ignores the absolute
value of $\psi $\ prepared by Bob. This shows that two photons are
not sufficient for a perfect eavesdropping with
$\mathcal{M}^\prime$. Yet the complete attack can be accomplished
with an additional photon $p_{3}$: Eve should store $p_{3}$,
execute $\mathcal{M^\prime}$, and eventually encode $p_{3}$
according to the conclusive outcome of $\mathcal{M^\prime}$; the
photon prepared in this way can be forwarded to Bob without risk
of detection.

The above analysis establishes that a perfect (i.e. with zero
QBER) eavesdropping can be realized with at least three photons in
a pulse. It also establishes that the measurement $\mathcal{M}$
represents a more powerful resource for Eve than
$\mathcal{M^\prime}$, for a number of reasons: it gives
information on the complete polarization state $\psi$ of the
photons, not only on Alice's operation; the probability of
conclusive results is $1/2$ rather than $1/4$; Eve knows about the
conclusiveness of her measurement immediately, rather than after
Alice's encoding, and can use this information to improve her
strategy. For these reasons hereafter we only study the robustness
of the scheme against the PNS$_{\mathcal{M}}$ attacks. We do it
following L\"{u}tkenhaus's approach in \cite{Lut00}.

Bob prepares photons with a phase-averaged weak-pulse laser; the
statistics of the photons in each pulse is described by
Eq.(\ref{Pn(miu)}). Given a forward-and-backward lossy channel
with transmissivity $t_{link}$, and with reference to the MM runs
of the protocol, we see that the encoded photons are revealed by
Bob's APDs with average probability:
\begin{equation}\label{gain}
p_{av}^{sign}=1-e^{-\mu\eta_{B}t_{link}}\simeq\mu\eta_{B}t_{link},
\end{equation}
where the approximation is valid for small values of the exponent.
$\eta_{B}$ is the quantum efficiency of Bob's detectors;
$t_{link}=10^{-(\alpha l+\Gamma_{c})  /10}$ is the transmissivity
of the channel, where $\alpha$ is the absorption
coefficient~\cite{foot2}, $l$ is the distance (in Km) between the
place in which the photon is prepared and the place in which it is
detected, and $\Gamma_{c}$ is a constant total loss-rate given by
Alice's encoding equipment and Bob's measuring apparatus. To the
signal revealed by Bob contribute also the dark counts per gating
windows, $d_B$, from his (two) detectors: $p_{av}^{dark}=2d_{B}$,
for a total signal probability equal to
\begin{equation}
p_{av}=p_{av}^{sign}+p_{av}^{dark}-p_{av}^{sign}p_{av}^{dark},
\end{equation}
where the last term represents the probability of a coincidence
between a dark count and a true signal photon. Now we can write
the necessary condition for security against the
PNS$_{\mathcal{M}}$ attacks as:
\begin{equation}\label{yield}
  p_{av}  >\widetilde{P}=\frac{1}{2}P_{n=3}\left(  \mu\right)
        +P_{n>3} \left(  \mu\right)
    =1-\left(1+\mu+\frac{\mu^{2}}{2}+\frac{1}{2}\frac{\mu^{3}}{6}\right)
        e^{-\mu}
\end{equation}
In Eq.(\ref{yield}) we conservatively assumed that the probability
of a conclusive outcome from $\mathcal{M}$ for more than three
photons in a single pulse is $1$. The meaning of the above formula
is that when the loss-rate is too high the probability to detect a
signal photon becomes smaller and smaller, eventually letting Eve
conceal under the expected losses.
\begin{figure}[h!]
  % Requires \usepackage{graphicx}
  \centering
  \includegraphics[width=10cm]{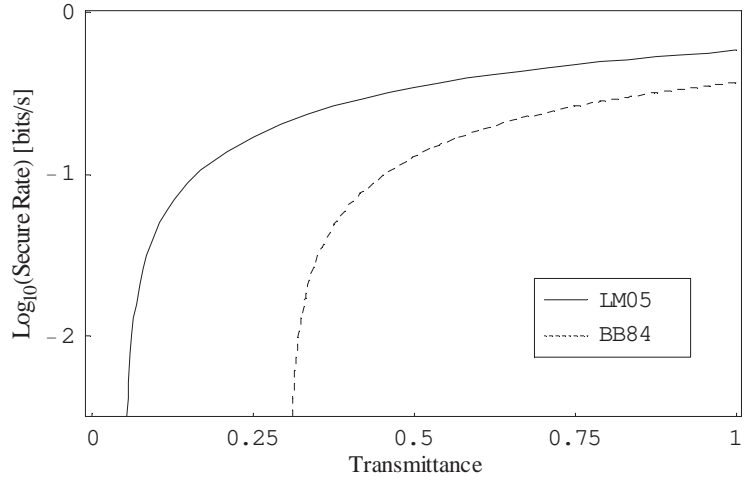}\\
  \caption{Secure rate versus transmittance for the protocols LM05
  (continuous line) and BB84 (dotted line) with the following
  parameters:
  $\mu=1$, $\eta_{B}=1$, $\Gamma_{c}=0$. The `secure rate'
  is defined by $(p_{av}^{sign}-\widetilde{P})$ for LM05 and by $(p_{av}^{sign}-P^*)$
  for BB84. See text for the explicit expressions of $p_{av}^{sign}$, $\widetilde{P}$ and $P^*$.}
  \label{FIG1}
\end{figure}
In Fig.\ref{FIG1} we plotted the logarithm of the difference
$(p_{av}^{sign}-\widetilde{P})$, that defines the security region
of LM05, versus the transmittance of the channel $t_{link}$, after
setting $\mu=1$, $\eta _{B}=1$ and $\Gamma_{c}=0$. We also plotted
the analogous curve for BB84 under the same settings. The purpose
is to show for which values of $t_{link}$ the two protocols are
secure against PNS attacks. It can be noted that despite the quite
high value of $\mu$ the security of LM05 is attained for almost
all the values of $t_{link}$. In order to reduce the probability
on the right side of Eq.(\ref{yield}) and increase the security of
the scheme we could decrease the value of $\mu$; however in this
way also the probability to detect a signal (Eq.(\ref{gain})) will
decrease. So there must be a tradeoff between these two opposite
requirements (security and signal rate) that defines an optimality
region for the scheme.

The tradeoff can be studied through the `gain of secure bits'
defined in \cite{Lut00}, that we rewrite here for LM05 protocol:
\begin{equation}\label{tau'}
    G_{sec}=p_{av}\left[ \beta(1-\tau^{\prime})-
                         f_{casc}h\left(e\right) \right]
            \qquad\quad{\text ;}\qquad\quad
    \tau^{\prime}=\tau\left(  e/\beta\right)\,.
\end{equation}
It represents the fraction of secure bits that can be distilled
from the transmitted bits after the procedures of error
correction~\cite{Ben92c,Bra94} and privacy
amplification~\cite{Ben95}.
$\beta=(p_{av}-\widetilde{P})/p_{av}>0$ is the security parameter:
until it is positive the protocol is secure against
PNS$_{\mathcal{M}}$ attacks. $f_{casc}$ is a function defined
in~\cite{Bra94} that takes into account the imperfect (although
efficient) error correction procedure performed with the Cascade
protocol; $h\left(  e\right)  $ is the Shannon entropy for the
QBER $e$; $\tau$ is the fraction of the error-corrected key which
has to be discarded during privacy amplification when only
single-photon pulses are taken into account~\cite{Lut99}; it is a
function of the QBER and amounts to~\cite{foot3}: $\tau(e)
=\log_{2}(1+4e-4e^{2})$ for $0\leq e\leq1/2$ and $\tau(e) =1$ for
$1/2<e\leq1$. Finally $\tau^{\prime}$ in Eq.(\ref{tau'})
represents the fraction of bits to discard after taking into
account multiphoton pulses: it amounts to $\tau$ scaled with the
security parameter $\beta$. The QBER $e$ is given by the
experiment according to the following expression
$e=(n_{err}+n_{D}/2)/n_{tot}$, where $n_{err}$ is the number of
error bits in the sifted key, $n_{D}$\ is the number of
`ambiguous' double clicks in Bob's APDs and $n_{tot}$ is the total
number of used bits. The importance of $n_{D}$ is theoretical:
usually $n_{D}\ll n_{err}$, and it can be completely neglected.
\begin{figure}[h!]
  % Requires \usepackage{graphicx}
  \centering
  \includegraphics[width=12cm]{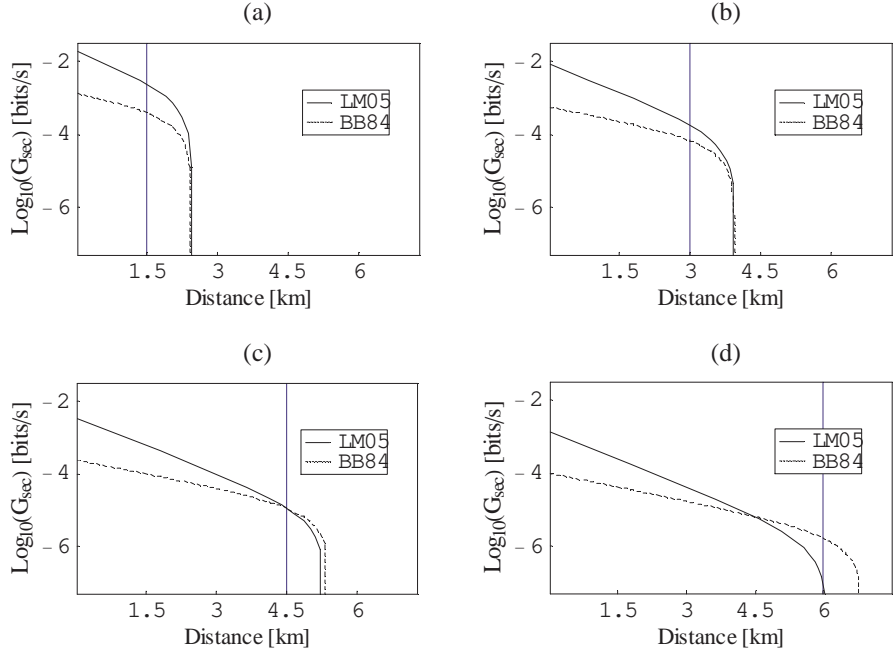}\\
  \caption{Secure rate vs Alice-Bob distance given the PNS attacks described in the
text. $\lambda=830$~nm, $\alpha=2.5$~dB/Km, $\Gamma_c=8$~dB,
$d_B=5\times10^{-8}$~counts/slot, $\eta_{B}=0.5.$} \label{FIG2a}
\end{figure}

In Fig.\ref{FIG2a} the gain $G_{sec}$ for both LM05 (continuous
lines) and BB84 (dotted lines) is plotted as a function of the
distance between Alice and Bob. We note that when Alice-Bob
distance is $l$ the total distance between the creation of the
photon and its final detection is $l$ for the BB84, and $2l$ for
the LM05, due to the double usage of the quantum channel. The BB84
implementation we adopted for comparison with LM05 is the one
reported in Ref.~\cite{Tow98} for the first optical fiber
communication window at wavelength around $0.8\mu{m}$. It is
worthwhile noting that the secure gain has a maximum in $\mu$ for
every fixed length $l$. Hence the pictures in Fig.\ref{FIG2a} have
been obtained by fixing four values of $l$ ($l_{1}=1.5$,
$l_{2}=3$, $l_{3}=4.5$, $l_{4}=6$ Km respectively plot \textit{a},
\textit{b}, \textit{c}, \textit{d}) both for LM05 and BB84, and
finding the values $\mu_{i}$ that provide a maximum for
$G_{sec}\left( \mu_{i}|l_{i}\right)$. We allowed $\mu_{i}$ to be
different in BB84 and LM05. Vertical lines have been drawn at the
typical distances $l_{i}$ ($i=1,...4$). The maximum distance for
both the protocols is between 6 and 7 Km for the parameters given
in the caption of Fig.\ref{FIG2a}. It can be seen that in
correspondence of the vertical lines $l_{1}$, $l_{2}$, $l_{3}$
(plots \textit{a}, \textit{b}, \textit{c}) the LM05 curves are
above the BB84 curves, while it is the opposite for $l_{4}$ (plot
\textit{d}). This means that for almost all the relevant distances
between Alice and Bob, the LM05 allows for a better gain of secure
bits, that directly reflects in higher distribution rates of
secure bits between the users. The improvement on small and medium
distances has two reasons: the first is the deterministic nature
of the protocol, that doubles the rate by removing the basis
reconciliation procedure; this reflects in a factor 2 for
$G_{sec}$ pertaining to LM05 respect to that pertaining to
BB84~\cite{Lut00}. The second reason is the two-way channel, that
provides the probability $\widetilde{P}$ of Eq.(\ref{yield})
considerably smaller than the analogous of BB84, given by
$P^*=P_{n\geq2}\left( \mu\right) =1-\left( 1+\mu\right) e^{-\mu}$.
It should be noticed that the same double channel also implies a
higher total loss-rate for LM05; then the increased gain of secure
bits is a non trivial result.

\section{Experiment}

\begin{figure}[h!]
  \centering
  \includegraphics[width=7cm]{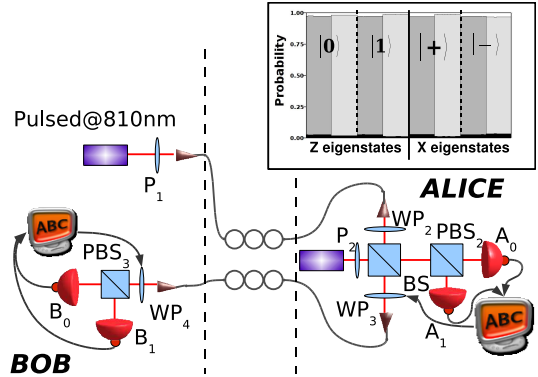}\\
  \caption{Experimental setup (see text for details).
  {\it Inset --} a typical communication test for different sets of
preparation by Bob and operation by Alice. Bob's preparation is
reported on the overlay ($\widehat{\sigma}_{z}$ and
$\widehat{\sigma}_{x}$ eigenstates). Alice's encoding is
represented by the lighter gray area for $I$ (logical `0'), and
darker gray for $i\widehat{\sigma}_{y}$ (logical `1'). The black
area represents the distribution of the QBER.} \label{setup}
\end{figure}

The experimental test of LM05 for QKD is realized exploiting
non-orthogonal polarization states of near infrared photons (see
Fig.\ref{setup}).

The photon source is a pulsed diode laser (Picoquant PDL 808) at
$810$nm with a repetition rate of 20MHz, pulse width $88$ps FWHM.
A pulse generator is used as sync source for the laser diode and a
detection circuit. The light pulses are first split in two, one
half goes at Bob's side for the initial state preparation for both
CM and MM runs, the other half is sent to Alice for the CM runs.

The first stage of the protocol is the preparation at Bob's side
of the qubits, encoded using a $\lambda/2$ waveplate
($\text{P}_1$), in four polarization linear states of the light
pulses attenuated to an average number of photons per pulse of
$\mu=(0.118\pm0.002)$. The prepared photons are launched into $5m$
long single mode fibers at $810$nm (Thorlabs P1-830A-FC)
connecting Alice and Bob. Before every test the fiber was aligned
using polarization control pads (Thorlabs FPC-560) so that any
polarization input state exits almost unchanged [the fibers proved
to remain stable for quite long periods ($\sim4$h), enough for
several runs after the alignment].
The second stage is at Alice's side. The switch between CM and MM
is passively realized via a 50/50 BS. \emph{Control Mode}: The
photons are polarization analyzed by a set composed by a
$\lambda/2$-waveplate ($\text{WP}_2$), a polarizing BS
($\text{PBS}_2$) and two APD (PerkinElmer SPCM-AQR-13-FC) modules
with quantum efficiency $\eta_B\sim50\%$ at $810$nm and dark
counts $\sim 300$cps ($\text{A}_0$ and $\text{A}_1$). The counts
rates are measured in a 8ns time window triggered by the sync
source, giving a dark counts per gating windows
$d_B\sim2.4\times10^{-6}$. To complete the control mode, Alice
injects in the BS, used for switching from CM to MM, a light pulse
generated by the second half of the pulse originated from the
diode laser. The pulses are polarization encoded in similar
fashion as at Bob's side ($\text{P}_2$), and attenuated to a mean
photon number per pulse almost 1/20 of the MM one~\cite{foot4}. A
couple of $\lambda/2$ waveplates ($\rm{WP}_{2,3}$) are used to
realize the $I$ and the $i\widehat{\sigma}_{y}$ operators
necessary for the \emph{Message Mode}~\cite{Cere06}. As last step,
the photon travels back to Bob through a different fiber, $5$m
long too, with a polarization control pads.

The photons coming from Alice are eventually polarization-analyzed
at Bob's side by $\rm{PBS}_3$ and a $\lambda/2$ waveplate
($\rm{WP}_4)$ set so that the photons are measured in the same
basis as they were prepared. The photons are collected after the
$\rm{PBS}_3$ into two multimode fibers and then detected by two
APD modules, $B_0$ and $B_1$. Counts out of $B_0$ and $B_1$ in a
8ns time window triggered to the sync source, can be associated to
logical values `0' and `1' corresponding to Alice encoding in the
MM runs. A typical result of a communication test is reported in
the inset of~Fig.\ref{setup} for different state preparations
performed by Bob and different encodings by Alice (all the eight
configurations of interest).

In our experimental tests we estimated the total QBER as
$e=\widetilde{n}_{err}/n_{tot}$, where $n_{tot}$ is the total
number of counts and $\widetilde{n}_{err}$ the counts in the
`wrong' detector. The best value we obtained for $e$ is
$(0.0248\pm0.0001)$. We have estimated a probability of
`ambiguous' double clicks $n_{D}/n_{tot}\sim9\times 10^{-4}$, a
factor $\sim 30$ lower than the probability of error bits $e$,
i.e. we can approximate $\widetilde{n}_{error}\sim n_{err}$ (see
discussion before Fig.~\ref{FIG2a}). The channel transmissivity is
estimated to be $t_{link}\sim0.27$ giving $\Gamma_{c}\sim5.7$dB.
These parameters allow to estimate the `secure bits gain',
$G_{sec}$, for LM05 and BB84 to be $0.018$ and $0.006$,
respectively. With a $20$MHz repetition rate laser this entails
the possibility to distribute secret bits with LM05 at $\sim360$
kbits/s, 3 times higher than BB84 ($\sim120$ kbits/s). The mean
photon number used in the experiment represents the optimal $\mu$
for distances up to $\sim3$ Km.

\section{Conclusion}

Our study shows the security of the LM05 protocol against a class
of PNS attacks, based on imperfections of Alice and Bob's
equipment. As a byproduct we found that LM05 allows for higher
distribution rates of secure bits respect to the BB84, for almost
all the relevant distances between Alice and Bob. In our analysis
we made the implicit assumption that Eve is clever enough not to
alter the statistics of the losses counted by Alice and
Bob~\cite{Lut02}. This means that in the frame of
PNS$_{\mathcal{M}}$ attacks Eve should distribute her `blocking
action' on both the paths (to and fro) between Alice and Bob,
otherwise resulting more easily detectable.

Furthermore we have reported on the first experimental test of
LM05 implemented with weak coherent state at $0.8\mu$m. We have
measured a QBER $e\sim0.024$ for a communication distance of $5$m,
which for the parameters of our setup allows for a secure bit rate
$\sim3$ times higher than BB84 for distances up to $\sim3$ Km.

\medskip

This work has been supported by the European Commission through
the Integrated Project `Scalable Quantum Computing with Light and
Atoms' (SCALA), Contract No 015714, `Qubit Applications' (QAP),
Contract No 015848, funded by the IST directorate, and the
Ministero della Istruzione, dell'Universit\'a e della Ricerca
((FIRB-RBAU01L5AZ and PRIN-2005024254)).


\begin{thebibliography}{99}
\bibitem{foot1} We assume that no phase reference on the prepared
pulses is available to Eve, so that she can only see the pulses
distributed according to Eq.(\ref{Pn(miu)}); see \cite{Lut00} and
\cite{Bra00}.
\bibitem{foot2} We consider for simplicity the same absorption coefficient in both
the forward and backward channels.
\bibitem{foot4} The difference in the photon mean number between CM runs and MM
runs can be positively exploited by Alice and Bob to further
increase the security of LM05 against PNS attacks. This result
descends from a straightforward application of the so called
`decoy state' technique~\cite{Hwa03,Wan05,Lo05,Zha06}, and will be
investigated in a future work.
\end{thebibliography}
\end{document}